\documentclass[12pt]{article}
\usepackage{graphicx}
\usepackage{color}
\usepackage{amsmath}
\usepackage{amssymb}
\usepackage{amscd}
\usepackage{amsthm}
\usepackage{amsopn}
\usepackage{xspace}
\usepackage{verbatim}
\usepackage{amsmath}
\usepackage{amsfonts}
\usepackage{epic}
\usepackage{eepic}
\usepackage{stmaryrd}
\usepackage{empheq}
\usepackage[active]{srcltx}
\usepackage[a4paper,margin=3cm]{geometry}
\definecolor{red}{rgb}{1,0,0}
\definecolor{green}{rgb}{0,1,0}
\definecolor{SeaGreen}{RGB}{46,139,87}
\definecolor{Maroon}{RGB}{128,0,0}

\newcommand{\R}{{\mathbb{R}}}

\def\PP{\mathcal P}

\renewcommand {\Re}{{\rm Re\,}}
\renewcommand{\Im}{{\rm Im\,}}
\newcommand {\pa}{\partial}

\newcommand {\ar}{\to}

\def\Ab{{\mathbf A}}

\def\0{\mathbf  0}

\def\XXint#1#2#3{{\setbox0=\hbox{$#1{#2#3}{\int}$ }
\vcenter{\hbox{$#2#3$ }}\kern-.6\wd0}}

 

\errorcontextlines=0 \numberwithin{equation}{section}
\theoremstyle{plain}
\newtheorem{theorem}{Theorem}[section]
\newtheorem{lemma}[theorem]{Lemma}

\newtheorem{proposition}[theorem]{Proposition}

\newtheorem{definition}[theorem]{Definition}
\newtheorem{remark}[theorem]{Remark}
\newtheorem{corollary}[theorem]{Corollary}

\title{On the domain of
  a  magnetic Schr\"odinger operator with complex electric potential.} 

\author{ 
  B. Helffer, Laboratoire de Math\'ematiques Jean Leray, \\CNRS and Universit\'e de Nantes, \\
  2 rue de la Houssini\`ere, 44322 Nantes Cedex France.\\
  and\\
  J. Nourrigat, 
LMR EA 4535 and FR CNRS 3399, \\Universit\'e de Reims Champagne-Ardenne,\\ Moulin de la Housse, BP 1039,
51687 REIMS Cedex 2, France.}

\date{}

\begin{document}
\bibliographystyle{siam}

\maketitle
\begin{abstract}
The aim of this paper is to review and compare the spectral properties of (the closed extension of )  $-\Delta + U$ ($V\geq 0$) and $-\Delta + i V$ in $L^2(\mathbb R^d)$ for $C^\infty$ real  potentials $U$ or  $V$ with polynomial behavior. 
The case with magnetic field will be also considered. More precisely, we would like to present the existing criteria for:
\begin{itemize}
\item essential selfadjointness or maximal accretivity
\item Compactness of the resolvent.
\item Maximal inequalities, i.e. the existence of $C>0$ such that, $\forall u\in C_0^\infty(\mathbb R^d)$,
\begin{equation}
||u||^2_{H^2(\mathbb R^d)} + || U  u||_{L^2(\mathbb R^d)}^2 \leq C \left( || (-\Delta + U) u||_{L^2(\mathbb R^d)}^2 + ||u||_{L^2(\mathbb R^d)}^2\right),
\end{equation}
or
\begin{equation}
||u||^2_{H^2} + || V u||^2 \leq C \left( || (-\Delta + i V) u||^2 + ||u||^2\right).
\end{equation}
 \end{itemize}
 Motivated by recent works with X. Pan, Y. Almog and D. Grebenkov (\cite{AHP, H, AH, AGH}), we will actually improve the known results in the case with purely imaginary potential.
 
\end{abstract}

\section{Introduction}
\label{sec:1}
In this paper, we  review and compare the spectral properties of (the closed extension of )  $-\Delta + U $ ($U \geq 0$) and $-\Delta + i V$ in $L^2(\mathbb R^d)$ and more precisely the criteria for:
\begin{itemize}
\item essential selfadjointness or maximal accretivity
\item Compactness of the resolvent.
\item Maximal inequalities, i.e. the existence of $C>0$ such that, $ \forall u\in C_0^\infty(\mathbb R^d)$,
\begin{equation}
||u||^2_{H^2(\mathbb R^d)} + || U u||_{L^2(\mathbb R^d)}^2 \leq C \left( || (-\Delta + U ) u||_{L^2(\mathbb R^d)}^2 + ||u||_{L^2(\mathbb R^d)}^2\right),
\end{equation}
or
\begin{equation}
||u||^2_{H^2} + || V u||^2 \leq C \left( || (-\Delta + i V) u||^2 + ||u||^2\right).
\end{equation}
 \end{itemize}
We will also discuss the magnetic case. In this case the operator reads:
$$
P_{\Ab,W} =  -\Delta_A + V := \sum_{j=1}^d (D_{x_j} - A_j(x))^2 + W (x)\,,
$$
where $\Ab= (A_1,\dots, A_d)$ is a $C^\infty$ vector fields on $\mathbb R^d$ 
and the maximal regularity is expressed in terms of the magnetic Sobolev spaces:
\begin{equation}\label{maxextmag}
\begin{array}{l}
|| (D-\Ab) u||^2_{L^2(\mathbb R^d,\mathbb C^d)} + \sum_{j,\ell} ||(D_j-A_j)(D_\ell-A_\ell) u||^2_{L^2(\mathbb R^d)} + || \,|W| u||_{L^2(\mathbb R^d)}^2 \\ \quad\quad \leq C \left( || P_{\Ab,W} u||_{L^2(\mathbb R^d)}^2 + ||u||_{L^2(\mathbb R^d)}^2\right),
\end{array}
\end{equation}

The question of analyzing $-\Delta + i V$ or more generally $P_{\Ab,i V}:=-\Delta_A + i V$ appears in many situations \cite{AH, AHP, AGH}. It seems therefore useful to present in a unified way, what is known on the subject  in the selfadjoint case and try to go further in the accretive case, where much less is known. 
 If we assume that the potential $V$ is $C^\infty$, we know that the operator is essentially selfadjoint starting from $C_0^\infty(\mathbb R^d)$ in the first situation and maximally accretive in the second case. Hence in the two cases the closed operator in consideration is uniquely defined by its restriction to $C_0^\infty$.\\
 At least for the selfadjoint case, the subject has a long story, in which T. Kato and his school plays an important role. We refer to \cite{SimSurvey} for a rather complete presentation with an exhaustive list of reference. One should also mention the  work of Avron-Herbst-Simon (1978)  \cite{AHS} which popularizes the basic questions on the subject and in particular the magnetic bottles.  \\
For the compactness of the resolvent, outside the easy case when $U  \rightarrow +\infty$, the story starts around the eighties  with the treatment of  instructive examples (Simon  \cite{Sim}, Robert \cite{Ro}) and in the case with magnetic field \cite{AHS} (the simplest example being for $d=2$ and $U=0$, when $B(x)\rightarrow +\infty$). In the polynomial case, many results are deduced as a byproduct of the analysis in Helffer-Nourrigat \cite{HN}, at least in the case when $V$ is a sum of square of polynomials. 
Using Kohn's type inequality, B.~Helffer and A. Morame (Mohamed) \cite{HM} (1988)  obtain more general results which can be combined with the analysis of A. Iwatsuka \cite{Iw} (1986). Another family of results using the notion of capacity can be found in \cite{KS,KMS} (see references therein).\\

T. Kato proves for example the inequality
\begin{equation}
|| \Delta u||_{L^1} + || U u||_{L^1} \leq  3 \,  || (-\Delta + U) u ||_{L^1}\,,\, \forall u \in C_0^\infty (\mathbb R^d)\,,\,
\end{equation}
under the condition that $U\geq 0$ and $U \in L^1_{loc}$.\\
The generalization to the $L^p$ ($p>1$) is only possible under stronger conditions on $U$. We will mention some of these results but will focus  on the $L^2$ estimates which are some times easier to obtain. In the case, when
 $U(x)=\sum_\ell  U_\ell (x)^2$, the maximal $L^2$ estimate is obtained as a byproduct of the analysis of the hypoellipticity (see H\"ormander \cite{Ho}, Rothschild-Stein \cite{RoSt} and  the book Helffer-Nourrigat  \cite{HNo} (including polynomial magnetic potentials)). This was then generalized to the case when $V$  is a positive polynomial by J. Nourrigat in an unpublished paper \cite{No1} circulating in the nineties and  used in the PHD of D. Guibourg \cite{Gu1, Gu2} defended in 1992, which considers the case when the electric potential  $U \geq 0$ and the magnetic potential ${\bf A}$ are polynomials (one chapter treats a more general situation). In his thesis Zhong (1993) proves the same result by showing that $\nabla^2 (-\Delta + U)^{-1}$ is a Calderon-Zygmund operator. Z.~Shen (1995) \cite{Sh}  generalizes the result to the case when $U$ is in the reverse H\"older class $RH_q$ ($q\geq \frac d 2$), a class which contains the positive polynomials.  \begin{definition}
 A locally $L^q$  function $\omega$ and strongly positive almost everywhere belongs to $RH_q$ if there
  exists a constant $C>0$ such that for any cube $Q$ in $\mathbb R^d$
  $$
  \left( \frac{1}{|Q|} \int \omega^q\, dx\right)^\frac 1q \leq C \left(  \frac{1}{|Q|} \int \omega\, dx\right)\,.
  $$
\end{definition}
One should also mention the unpublished  thesis of Mba-Y\'eb\'e \cite{MY} defended in 1995. Together with the techniques developed by Guibourg, some of  his techniques are  useful for the improvments presented in the last section.\\

  Z. Shen considers also the case with magnetic fields in 1996 \cite{Sh2}. Further progress are obtained in the thesis of B. Ben Ali \cite{BA} (2007), 
   published in \cite{ABA}, \cite{BA1} and  \cite{BA2}.  The methods applied by Shen and Auscher--Ben Ali include the Fefferman--Phong inequalities, the Calder{\'o}n--Zygmund decompositions, and various techniques of interpolation. We will come back to one of these results in Subsection \ref{ss2.2}.

\section{Kohn's approach}\label{s2}
This approach was mainly used for getting the compactness of the resolvent. Except in a few cases, these estimates do not lead to the maximal regularity but are enough for getting the compactness and we will see that surprisingly they could also be a step for proving $L^2$-maximal estimates.
\subsection{Self-adjoint case}\label{ss2.1}
Here we mainly refer to \cite{HM} (see also \cite{Mef}, \cite{HN}). 
We analyze the problem for the family of operators~:
\begin{equation}
\PP_{\Ab, V}  = \sum_{j=1}^d  (D_{x_j}- A_j(x))^2 + \sum_{\ell=1}^p  U_\ell (x)^2\;.
\end{equation}
Here  the magnetic potential
 $\Ab (x)=(A_1(x), A_2 (x),\cdots, A_n(x))$ is supposed to be $C^\infty$
 and  that $U_j\in C^\infty$. Under these conditions, the operator is 
\index{self-adjoint!essentially--}essentially self-adjoint on $C_0^\infty (\mathbb R^d)$. We note also that it has the form~:
$$
\PP_{\Ab, V}   = \sum_{j=1}^{d+p}  X _j^2 = \sum_{j=1}^d X_j^2 + \sum_{\ell =1}^p Y_\ell^2\;,
$$
with
$$
X_j = (D_{x_j} - A_j(x))\,,\, j=1,\dots,d\;,\; Y_\ell = U_\ell\;,\; 
\ell =1, \dots,  p\;.
$$
In particular, the magnetic field 
 is recovered by observing that
$$
B_{jk} = \frac 1i [X_j, X_k] = \pa_j A_k - \pa_k A_j \;,\;\mbox{ for } j,k = 1,\dots,d\;.
$$
Of course, when $U \ar + \infty$, it is
well known that the operator has 
a compact resolvent.(see the argument below).\\
On the opposite, when $V=0$ and  $d=2$ if $B(x)=B_{12} \geq 0\,$,  one immediately deduces from the trivial inequality:
\begin{equation}
\int B(x) |u(x)|^2 dx \leq ||X_1u||^2 + || X_2 u ||^2   = \langle \PP_{\Ab, V}  u\;|\; u \rangle\;.
\end{equation}
that  $\lim_{|x| \to +\infty}B(x)= + \infty$ implies that the operator has a  compact resolvent.  A typical example is 
$$
A_1(x_1,x_2)= - x_2 x_1^2\;,\; A_2(x_1,x_2) = + x_1 x_2^2\;.
$$
In order to treat more general situations, we introduce the quantities:
\begin{equation}
\label{eq.mqx}
\check m_q(x) = \sum_\ell  \sum_{|\alpha|=q} |\pa_x^\alpha U_\ell|
+ \sum_{j<k}  \sum_{|\alpha|=q-1} |\pa_x^\alpha B_{jk} (x)|\;.
\end{equation}
It is easy to reinterpret this quantity in  terms of 
commutators of the $X_j$'s.\\
When $q=0$, the convention is that
\begin{equation}
\check m_0(x) = \sum_\ell |U_\ell(x) |\;.
\end{equation}
Let us also introduce
\begin{equation}
\check m^r (x) = 1 + \sum_{q=0}^r \check m_q(x)\;.
\end{equation}
Then the criterion proven by Helffer-Mohamed in 1988 (\cite{HM})  is
\begin{theorem} \label{thcompres}~\\
Let us assume that there exist $r$ and $C$ such that
\begin{equation}
\label{eq.mr1mr}
\check m_{r+1}(x) \leq C\; \check m^r(x)\;,\;\forall x\in \mathbb R^d\;,
\end{equation}
and
\begin{equation}
\check m^r(x) \ar + \infty\;,\; \mbox{ as } |x| \ar + \infty\;.
\end{equation}
Then, with $U=\sum_\ell U_\ell^2$,  $\PP_{\Ab, U}(h)$ has a  compact resolvent\index{compact!resolvent}.
\end{theorem}

\begin{remark}~\\
It is shown in  \cite{Mef}, that one can get the same result as in Theorem \ref{thcompres}  under
 the weaker assumption that
\begin{equation}\label{Meft}
\check m_{r+1}(x) \leq C \, [\check m^{r}(x)] ^{1+ \delta}\;,
\end{equation}
where $\delta= \frac{1}{2^{r+1} -3}$ ($r\geq 1$).  This result is optimal for $r=1$ according to a counterexample by A.~Iwatsuka \cite{Iw} who  has exhibited 
 an example of a Schr{\"o}dinger operator which has a non compact resolvent
 and such that $\sum_{j<k}|\nabla B_{jk} (x)|$ has the same order as 
 $\sum_{j<k} |B_{jk}|^2$.\\
Other generalizations are given in \cite{Sh} (Corollary 0.11) (see also references therein  and \cite{KS} for a quite recent contribution including  other references).\\
One can for example replace $\sum_j U_j^2$ by a more general  $U$ and the conditions on the $m_j$'s  can be reformulated in terms of the variation of $U$ and $B$ in suitable balls. In particular A.~Iwatsuka \cite{Iw} showed that a necessary condition
 is~:
\begin{equation}
\int_{B(x,1)} \left(U(x) + \sum_{j<k} B_{jk}(x)^2\right) dx \ar +\infty \mbox{ as } |x|\ar +\infty\;,
\end{equation}
where $B(x,1)$ is the ball of radius $1$ centered at $x$.
\end{remark}
\subsection{The accretive case : maximal accretivness}
There is  a general statement  (see for example  \cite{HN}, \cite{AHP})
about the maximal accretiveness of $\PP_{\Ab,W}:= -\Delta_A + W$, when $U \geq0$.
\begin{theorem}\label{Theorem2.1}
Consider the magnetic Schr\"odinger operator $\PP:= \PP_{\Ab,W}$ defined  on $\R^d$
with $\Ab\in C^\infty(\mathbb R^d,\mathbb R^d)$ and $W=U+i V \in
C^\infty(\mathbb R^d,\mathbb C)$ such that
\begin{equation}\label{2.9}
U (x) \geq 0\,.
\end{equation}\label{2.10}
Then the operator $\PP$ is  maximally accretive. Moreover
 \begin{equation}
\PP_{\Ab,W} = (P_{\Ab,\bar W})^*\,.
\end{equation}
\end{theorem}
We extend  Theorem \ref{thcompres} to the family of operators:
\begin{equation}
P_{{\Ab},W} = \sum_{j=1}^d  (D_{x_j}- A_j(x))^2 + \sum_{\ell=1}^p  U_\ell
(x)^2 + i\, V(x)\;.
\end{equation}
Here $U= \sum_{\ell=1}^p  U_\ell
(x)^2 $ and      $V$   is  $C^\infty$. We note also that it has the form:
$$
P_{{\Ab},W}  = \sum_{j=1}^{d+p}  X _j^2 = \sum_{j=1}^d X_j^2 + \sum_{\ell
  =1}^p Y_\ell^2 + X_0\;,
$$
with
$$
X_j = (D_{x_j} - A_j(x))\,,\, j=1,\dots,d\;,\; Y_\ell = U_\ell\;,\; 
\ell =1, \dots,  p\;,\; X_0= i V\,.
$$
We introduce the new quantity:
\begin{equation}
\label{eq.mqx}
\check m_q(x) = \sum_\ell  \sum_{|\alpha|=q} |\pa_x^\alpha U_\ell|
+ \sum_{j<k}  \sum_{|\alpha|=q-1} |\pa_x^\alpha B_{jk} (x)| +
\sum_{|\alpha|= q-1} |\pa_x^\alpha V|\;.
\end{equation}
and keep for $\check m^r (x)$ the same definition as in previous subsection:
\begin{equation}
\check m^r (x) = 1 + \sum_{q=0}^r \check m_q(x)\;.
\end{equation}
Then the criterion reads
\begin{theorem} \label{thcompresnew}~\\
Let us assume that there exist $r$ and $C_0$ such that
\begin{equation}
\label{eq.mr1mr}
\check m_{r+1}(x) \leq C_0\; \check m^r(x)\;,\;\forall x\in \mathbb R ^d\,.
\end{equation}
In this case, we say that $(\Ab,W) \in \mathcal T (r,C_0)$.\\
Then there exist $\delta >0$ and $C_1:= C_1 (C_0)$ such that, $\forall u \in C_0^\infty (\mathbb R^d)$
\begin{equation}
|| (\check m^r(x))^\delta  u||^2 \leq C_1 \left( ||P_{{\Ab},W} u||^2 + ||u||^2 \right).
\end{equation}
\end{theorem}
\begin{remark}
The proof, which was first given in a particular case in \cite{AHP}, will show that we can take $\delta = 2^{-r}$ which is in general not optimal. This $\delta$ can indeed be improved when $U = 0$ and $\Ab=0$ (see \cite{AGH}) but this improvment will
 not be used in this paper.
 \end{remark}
\begin{corollary}
Under the same assumptions, if 
\begin{equation}
\check m^r(x) \ar + \infty\;,\; \mbox{ as } |x| \ar + \infty\;.
\end{equation}
Then $\PP_{{\Ab},W}(h)$ has a  compact resolvent.
\end{corollary}
Before entering into the core of the proof, we observe that we can replace $\check m^r(x)$ by an equivalent $C^\infty$ function $\Psi(x)$ which has the property 
that there exist constants $C_\alpha$ and $C >0$ such that~:
\begin{equation}\label{eqw}
\begin{array}{l}
\frac 1C \Psi(x) \leq  \check m^r (x) \leq C \Psi(x)\;,\\
| D_x^\alpha \Psi(x) | \leq C_\alpha \Psi (x)\;.
\end{array}
\end{equation}
Indeed, it suffices to replace quantities like 
$\sum |u_{k}|$  by $(\sum |u_{k}|^{2})^{1/2}$,
in the definition \eqref{eq.mqx} of
$\check m_{q}$. The second condition is a consequence of \eqref{eq.mr1mr}.\\
In the same spirit as in \index{Kohn's~proof} Kohn's proof, let  us introduce:
\begin{definition}~\\
 For all $s >0$, we denote by $M^s$ the space of  $C^\infty$ real functions $T$ such that there exists $C_s$ such that:
\begin{equation}
|| \Psi^{-1+s} T u ||^2  \leq C_s \left(|| P_{{\Ab},W} u|| \;|| u ||+ || u ||^2 \right)\;,\; \forall u \in C_0^\infty(\mathbb R^d )\;.
\end{equation}
\end{definition}
We observe that
\begin{equation}\label{obs1}
U_\ell \in M^1\;,
\end{equation}
and we will show the
\begin{lemma}\label{lemmecro1}
\begin{equation}
i \, [X_j,X_k] \in M^{\frac 12} \;,\; \forall j, k =1, \dots, d\;.
\end{equation}
and
\begin{equation}
V \in M^{\frac 12}
\end{equation}
\end{lemma}

\noindent Another claim is contained in the 
\begin{lemma}\label{lemmecro2}~\\
If $T$ is in $M^s$ and $|\pa_x  ^\alpha T|\leq C_\alpha \Psi$ then
$i [X_k,T] \in M^{\frac s2}$, when $|\alpha| =1$ or $|\alpha|=2\,$.
\end{lemma}
Assuming these two lemmas, then it is clear that 
\begin{equation*}
\Psi(x) \in M^{2^{-r}}\,.
\end{equation*}
Lemma \ref{lemmecro2}
and (\ref{obs1}) lead to
\begin{equation*}
\pa_x^\alpha U_\ell \in M^{2^{-|\alpha|}}\,,
\end{equation*}
and
we deduce from  Lemmas \ref{lemmecro1}  and \ref{lemmecro2}:
\begin{equation*}
\pa_x^\alpha B_{jk} \in M^{2^{-(|\alpha|+1)}}\,.
\end{equation*}
The proof of Theorem \ref{thcompresnew} then becomes easy.

\noindent{\bf Proof of Lemma \ref{lemmecro1}}~\\
We start from the identity (and observing that $X_j^*=X_j$)~:
\begin{equation*}
\begin{array}{ll}
|| \Psi^{- \frac 12} [X_j,X_k] u ||^2&= \langle (X_j X_k - X_k X_j)u \;|\;  \Psi^{-1}  [X_j,X_k] u\rangle \\
&= \langle X_k u \;|\; X_j \Psi^{-1} [X_j,X_k] u\rangle \\
&\quad  -  \langle X_j u \;|\;   X_k\Psi^{-1}  [X_j,X_k] u\rangle \\
&= \langle X_j u \;|\; \Psi^{-1} [X_k,X_j]X_k  u \rangle \\
& \quad  - \langle X_k u \;|\; \Psi^{-1} [X_k,X_j] X_k u \rangle \\
&\quad  + \langle X_ju \;|\; [X_k, \Psi^{-1} [X_k,X_j]]u \rangle \\
&\quad  - \langle X_k u \;|\; [X_j, \Psi^{-1} [X_k,X_j]]u \rangle\;.
\end{array}
\end{equation*}
If we observe that $\Psi^{-1} [X_k,X_j]$ and $[X_k ,\Psi^{-1} [X_k,X_j]  ]$ are bounded (look at the definition of $\Psi$),  we obtain~:
\begin{equation*}
|| \Psi^{- \frac 12} [X_j,X_k] u ||^2 \leq C \left( ||X_k u ||^2 + || X_j u ||^2
 + ||u||^2 \right)\;.
\end{equation*}
We just observe that
\begin{equation}
\sum_j ||X_j u||^2 = \Re\langle P_{{\Ab},W} u\,|\, u\rangle\,.
\end{equation}
This ends the proof of the first part of the lemma.\\
For the second part, we start from 
$$
\Im \left(\langle \Psi^{-1} P_{{\Ab},W}  u\,,\, Q u\rangle\right)
 = ||\psi^{-\frac 12} Qu ||^2 + \Im \left(\langle \Psi^{-1} \sum_{j}
 X_j^2  u\,,\, V u\rangle\right)\,,
$$
and observe that
$$
\langle \Psi^{-1} 
 X_j^2  u\,,\, V u\rangle
= \langle \Psi^{-1} V  X_j u \,\, X_j u\rangle
 + \langle [X_j, \Psi^{-1} V] X_j u\,|\, u\rangle\,.
$$
Hence 
$$
\Im \langle \Psi^{-1} 
 X_j^2  u\,,\, V u\rangle
= \Im  \langle [X_j, \Psi^{-1} V] X_j u\,|\, u\rangle\,.
$$
Then using the property of $V$ and $\Psi$, we get the proof easily.

\noindent{\bf Proof of Lemma \ref{lemmecro2}}~\\
Let $T \in M^s$. For each $k$, we can write~:
\begin{equation*}
\begin{array}{ll}
|| \Psi^{- 1+ \frac s2} [X_k,T] u ||^2&=
 \langle  \Psi^{-1+s} (X_k T -T X_k)u \;|\;  \Psi^{-1} [X_k,T] u \rangle \\
&=
 \langle  \Psi^{-1+s} X_k T u \;|\;  \Psi^{-1} [X_k,T] u \rangle \\
&\quad  - 
 \langle  \Psi^{-1+s} T X_k u \;|\;  \Psi^{-1} [X_k,T] u \rangle \\
& 
= \langle \Psi^{-1+s} Tu \;|\; \Psi^{-1}[X_k,T] X_k u\rangle \\
  &\quad - \langle X_k u \;|\; \Psi^{-1} [X_k,T] \Psi^{-1+s} Tu \rangle \\
& \quad + \langle  T u \;|\; [X_k, \Psi^{-2 +s} [X_k,T]] u \rangle  \\
& = \langle \Psi^{-1+s} Tu \;|\; \Psi^{-1}[X_k,T] X_k u\rangle \\
  &\quad - \langle X_k u \;|\; \Psi^{-1} [X_k,T] \Psi^{-1+s} Tu \rangle \\
& \quad + \langle \Psi^{-1+s} T u \;|\; \Psi^{1-s}[X_k, \Psi^{-2 +s}
 [X_k,T]] u \rangle\;. 
\end{array}
\end{equation*}
We now  observe, according to the assumptions of the lemma and the
 properties of $\Psi$,  that $\Psi^{1-s}[X_k, \Psi^{-2+s} [X_k,T]]$
 and $\Psi^{-1} 
 [X_k,T]$ are bounded.

So finally  we get~:
\begin{equation*}
|| \Psi^{- \frac 12} [X_k,T] u ||^2 \leq C \left ( ||\Psi^{-1 +s} Tu
   ||^2 + ||X_k u ||^2 + ||u||^2\right) \;. 
\end{equation*} 
This ends the proof of the lemma.

\subsection{Nourrigat-Guibourg-Shen results}\label{ss2.2}
The results where first obtained in the polynomial case ($V=0\,,\, U\geq 0$), then for ${\Ab}$ and $U  \geq 0$ satisfying a condition "\`a la Helffer-Mohamed" and then with condition of type Reverse-H\"older. We take a version presented in Shen \cite{Sh2}  as a consequence of his Theorem 0.9 (see  p. 820 lines -13  to -8) which in addition refers to \cite{Sh4}. We only write the $L^2$ criterion.

\begin{theorem}\label{thSh} For $d\geq 3$, let us assume that ${\bf A} \in C^{r+2}(\mathbb R^d)$, $U \in C^{r+2} (\mathbb R^d)$ and $U\geq 0$ and
\begin{equation}
\sum_{|\beta|=r+1} |\partial_x^\beta B(x)| + \sum_{|\beta|=r+2} |\partial_x^\beta U(x)| \leq C \, \check m_{B,U}(x)\,,
\end{equation}
where
\begin{equation}
\check m_{B,U}(x) =\sum_{|\beta| \leq r} |\partial_x^\beta B (x)| + \sum_{|\beta|\leq r+1} |\partial_x^\beta U(x)| +1\,.
\end{equation}
Then 
\begin{equation}
\sum_{1\leq j,\ell \leq d} || (D_{x_j}-A_j(x))(D_{x_\ell} -A_\ell (x)) u ||_2 \leq C \left( || P_{\Ab,U} u||_2 + ||u||_2\right)\,,\, \forall u \in C_0^\infty (\mathbb R^d)\,.
\end{equation}
\end{theorem}

\begin{remark}\label{shend2}
As observed in \cite{Sh2}, the statement appearing in the thesis of D. Guibourg (Chapter B) is a little weaker. When $d=2$, according to \cite{Sh3}  there is no particular
difficulty. The assumption $(RH)_{n/2}$ appearing in \cite{Sh2} needs to change
to $(RH)_p$ for some $p>1$, and the size estimate for the fundamental
solution should have a logarithm. Otherwise, all results should be true.
\end{remark}

\section{Nilpotent approach}\label{s3}
The basic idea is to start from the maximal hypoellipticity of the H\"ormander operator $\mathcal P:= \sum_j \check X_j ^2 + \check X_0$ where the $\check X_j$'s are real vector fields satisfying the so called H\"ormander condition.
When considering the special case of a stratified group $\mathcal G = \mathcal G_1\oplus \cdots \oplus \mathcal G_r$, with the $\check X_j$'s ($j=1,\dots,k$) being a basis of $\mathcal G_1$, $\check X_0 \in \mathcal G_2$ and 
 the $\check X_j$ ($j=0,\dots,k$) generating the Lie algebra of rank $r$. \\
 We get for any induced representation $\Pi$ of $\mathcal G$ in $\mathcal H_\Pi\sim L^2(\mathbb R^k)$  the maximal estimate
 \begin{equation}
|| \Pi (\check X_0) u ||_{\mathcal H_\Pi} + \sum_{j,\ell=1,\dots,k} || \Pi (\check X_j\, \check X_\ell) u ||_{\mathcal H_\Pi} \leq C ||\pi (\mathcal P) u||_{\mathcal H_\Pi} \,,\, \forall u \in \mathcal S_\Pi\,.
\end{equation}
We can then use a variant of Proposition 1.6.1 in \cite{HNo}, to get for any polynomial $V$ with $k$ variables the inequality
\begin{equation}
\sum_{j,\ell} || \pa^2_{x_jx_\ell}u|| + || V(x) u|| \leq C\, || (-\Delta + i V (x)) u ||\,,\, \forall u\in C_0^\infty (\mathbb R^k)\,.
\end{equation}
We can indeed find $\mathcal G$, a subalgebra $\mathcal V$ and $\ell\in \mathcal G^*$ such that $\Pi$ is unitary equivalent to the representation $\pi_{\ell,\mathcal V}$ with
$$
\pi_{\ell, \mathcal V} (\check X_j) =  \pa_{x_j} \mbox{ for } j=1,\dots,k\,,\, \pi_{\ell, \mathcal V} (\check X_0) =i \, V(x)\,.
$$

As observed a long time ago, the same approach but looking only at \break $U(x) =\sum_j U_j(x)^2$ and using  the maximal hypoellipticity of $\sum_j \check X_j^2$, we obtain
\begin{equation}
\sum_{j,\ell} || \pa^2_{x_jx_\ell}u|| + || U(x) u|| \leq C\, || (-\Delta +  U(x)) u ||\,,\, \forall u\in C_0^\infty (\mathbb R^d)\,.
\end{equation}

It is then natural to ask if it is true for any polynomial $U \geq 0$ and more generally if we can  relax the polynomial condition. Here we refer to Guibourg or Nourrigat\footnote{There are actually two different proofs proposed by J. Nourrigat a rather direct one and another based to the analysis of $\sum_j \check X_j^2 + i  \check X_0$ the difficulty (but this  was sometimes treated in \cite{HN}) that this operator is no more hypoelliptic.} \cite{No1, No2} for the first point (this is indeed true) and to Shen \cite{Sh}  for the second point.\\
There is indeed a more general class of reverse H\"older potentials:
\begin{definition}
We say that  a non negative $\omega$ belongs to the reverse H\"older class if there exists $C$ such that, $\forall x\in \mathbb R^d$, $\forall r>0$,
\begin{equation}
\sup_{y\in B(x,r)}  \omega (x)  \leq \frac{C}{|B(x,r)|}\, \int_{B(x,r)} \omega (y)\, dy\,.
\end{equation}
\end{definition}
Note that a non negative polynomial satisfies this condition.\\

 Then other authors work on the subject with the aim of obtaining $L^p$ estimates \cite{ABA,BA}. The case with magnetic fields is always considered.

\section{Maximal estimates for the complex Schr\"odinger operator with magnetic potentials (non necessarily polynomial case)}\label{s1}

\subsection{Main statement}
We consider as before $$W = \sum_\ell U_\ell^2 + i V\,,$$
and the associated complex Schr\"odinger operator $ P_{{\Ab},W} $.

 \begin{theorem}\label{t-princ} If   $(\Ab,W) \in \mathcal T (r,C_0)$,  there exists
 $C>0$ such that, for all  $u\in C^{\infty}_0 (\R^d)$:
   \begin{equation}
    \Vert |W| u \Vert ^2  \leq C\left(  \Vert P_{{\Ab},W} \, u \Vert ^2 + ||u||^2\right) \,.
    \end{equation}
    \end{theorem}

 \subsection{H\"ormander's metrics and partition of unity.}\label{s2}
We introduce, for $t\in [0,1]$ and $x\in \mathbb R^d$, 
\begin{equation}
\Phi (x , t)   = \sum_\ell  \sum_{|\alpha|\leq r} \,  t^{|\alpha |+1}\,  |\pa_x^\alpha U_\ell (x)|
+ \sum_{j<k}  \sum_{|\alpha|\leq r-1}  t^{|\alpha|+2} \, |\pa_x^\alpha B_{jk} (x)| +
\sum_{|\alpha|\leq r-1}  t^{|\alpha |+2} |\pa_x^\alpha V (x)|\;.
\end{equation}
 As in  Mba-Y\'eb\'e  \cite{MY}, we introduce a parameter  $\mu \geq 1$  to be determined later and which is at the moment arbitrary and we define:
$$ R(x , \mu) = \sup \{ t \in [0, 1] , \ \ \ \  \Phi (x , t) \leq \mu \} $$

  \begin{proposition}\label{VL-1} If $(\Ab,W) \in \mathcal T (r,C_0)$ (see condition (\ref{eq.mr1mr}), 
  there exists  $C_2>1$ such that, for all $t\in (0, 1)$,
  we have the implication:
  $$ |y-x|\leq t \Longrightarrow \Phi (y, t) \leq C_2 \Phi (x , t) + C_2 \, t^{r+1} \,. $$
  \end{proposition}

  {\it Proof.} For all $x$ et $u$ in $\R^d$ such that $|u| = 1$,
 for all  $t$ and  $\theta$ such that $0 < \theta \leq t \leq 1$, let us introduce:
  $$ \Psi (x , u, \theta , t) = \Phi (x + \theta  u,  t)\,.$$
  Using  Taylor's formula  with integral remainder, we can write, if  $\theta \leq t$\,,
  $$ \Psi (x , u, \theta , t) \leq C \,  \Phi (x , t) + C R(t)\,, $$
  with 
  $$ R(t) =  t^{r+1}\, \sum_\ell   \sum _{|\beta | = r+1 }
  \int _0 ^{\theta} | \pa_x^{\beta }  U_\ell ( x + \sigma u )| \,d\sigma + ...  $$
  $$ ... + C  t^{ r+ 1}  \sum_{j<k}  \sum _{|\beta | = r }   \int _0 ^{\theta} | \pa_x^{\beta} B_{jk} (x+ \sigma u )|\, d\sigma +  C  t^{ r + 1}
   \sum _{|\beta | = r }  \int _0 ^{\theta} | \pa_x^{\beta} V (x+ \sigma u )| d\sigma\,.  $$ 
  Using condition (\ref{eq.mr1mr}) and $\theta \leq t\leq 1\,$, there exists $C>0$ such that
  $$ R(t) \leq
  C \int _0 ^{\theta}  \Psi (x , u, \sigma , t)\, d\sigma + C t^{ r+1} \,. $$
 We now apply Gronwall's Lemma and obtain the existence of $C_2>0$ such that, for $\theta \leq t\leq 1$\,,
 $$
  \Psi (x , u, \theta , t) \leq C_2 \Phi (x,t)  + C_2  t^{r +1} \,.$$
  This achieves the proof of the proposition.\\

\begin{proposition}\label{VL} Let $C_2 >1$ the constant of
Proposition \ref{VL-1}.
 Then we have:
    $$ |y-x|\leq \frac {R(x , \mu)} {2 C_2}  \Longrightarrow
 \frac {1} { 2 C_2 } \leq \frac   { R(y , \mu)} { R(x , \mu)}
  \leq  2 C_2   $$
  \end{proposition}

    {\it Proof.}  We apply Proposition \ref{VL-1} with  $t_0 = R(x , \mu) \leq 1$.
    If $  |y-x|\leq R(x , \mu)  $, we have 
    $$  \Phi (y, t_0) \leq C_2 (  \Phi (x, t_0)
 + t_0^{r+1}) \leq 2 C_2 \mu\,.
 $$
  We have indeed
 $ \Phi (x, t_0) \leq \mu$ and 
 $  t_0^{r+1} \leq 1 \leq \mu$. \\
 Consequently $t_1 = t_0 / 2 C_2 =
 R(x , \mu)/ (2 C_2)$ satisfies  $t_1 \leq 1$  and  $ \Phi (y, t_1) \leq \mu$.
 Therefore we get  $t_1 \leq R(y, \mu)$,  hence the first inequality above. \\
  If now
 $  |y-x|\leq R(x , \mu) / (2 C_2) $, we deduce  $  |y-x|\leq R(y , \mu)$, and, permuting the roles of 
  $x$ and $y$, we effectively get  $$ R(y , \mu) \leq  2 C_2    R(x , \mu)\,.
  $$
 This achieves the proof of the proposition.\\

 This proposition shows that the metric defined on  $\R^d$ by
    $ g_x (t) = |t|^2 / R(x, \mu)^2$ ($x\in \R^d$, $t\in \R^d$),  is slowly varying in the sense of Definition 18.4.1 in \cite{Ho-2}. Moreover, the constant in the definition  can be chosen independently of $\mu$.  We deduce from Lemma~18.4.4 in  \cite{Ho-2}
    the following proposition.

     \begin{proposition}\label{PU} For any $\mu \geq 1$, there exist a sequence of real valued  functions
    $(\varphi_j)$ in  $C^{\infty}_0 (\R^d)$, and a sequence
    $(x_j)$ in $\R^d$, such that:
    \begin{itemize}
    \item
     \begin{equation}\label{phi-j-1} \sum _j \varphi _j(x)^2 = 1 \,,\, \forall  x\in \R^d\,. \end{equation}
    \item 
     \begin{equation}\label{phi-j-2} {\rm supp\,} \varphi_j \subset B(x_j , R(x_j ,\mu) )\,. \end{equation}
    \item For any  multi-index $\alpha $, there exists $\hat C_{\alpha } >0$, independent
    of $\mu$, such that
     \begin{equation}\label{phi-j-3} \sum _j |\partial ^{\alpha} \varphi _j(x)|^2 \leq
    \frac { \hat C_{\alpha } } { R(x , \mu) ^{2|\alpha |}}\,.  \end{equation}
   \item   There exists $\hat C>0$, independent of  $\mu$, such that, for  $k=1,2$, for any  $u$ in  $C^{\infty }_0(\R^d)$,
     \begin{equation}\label{phi-j-4}  \int _{\R^d} \frac {|u(x)|^{2} } {R(x , \mu)^{2k}} dx
    \leq \hat C \,   \sum _{j}
    \int  _{\R^d} \frac {\varphi_j (x)^2 |u(x)|^2 } {R(x_j , \mu )^{2k}} dx  \,.
    \end{equation}
    
    \end{itemize}
    \end{proposition}

   \subsection{Proof of Theorem \ref{t-princ}.}\label{s4}  
  Just observing that:
 $$ \Re < P_{U+iV} f, f > = ||(D-\Ab) f||^2 + \int U |f|^2 dx\,,
 $$ 
 we obtain:
  \begin{lemma}\label{lemm-interpol} For all  $f\in C^{\infty }_0 (\R^d)$,
 we have:
  \begin{equation}\label{interpol}  \sum _j  \Vert  (D_{x_j}- A_j) f \Vert ^2 + \sum _{\ell = 1 } ^p 
   \Vert   U_\ell  f \Vert ^2   \leq \Vert P_{{\Ab},W} f \Vert \ \Vert f \Vert \,. \end{equation} 
\end{lemma}

 \begin{proposition}\label{estim-loc} For any  $\mu >1$,  let  $(x_m)$   be a  sequence of points in $\R^d$ as in Proposition~\ref{PU}.  Let $(\Ab,W)\in \mathcal T (r,C_0)$. Then there exist $\mu_0 >1$ and $C_3$ (depending only on $r$ and    $C_0$)  such that, for any  $m$ such that  $R(x_m , \mu) \leq 1/2$,  and for any  $f\in C^{\infty} _0(\R^d)$ supported in the ball  $B_m = B(x_m , R(x_m, \mu))$ and $\mu \geq \mu_0$,  
 \begin{equation}\label{EL}  \frac {\mu ^{\delta } } { R(x_m , \mu)^2 } \Vert f \Vert
+  \frac {\mu ^{\delta /2 } } { R(x_m , \mu) }  \Vert (D -\Ab ) f \Vert
\leq C_3 \Vert P_{{\Ab},W}  f \Vert \end{equation}
where  $\delta $ is the constant given by Theorem \ref{thcompresnew}. 
 \end{proposition}

 {\it Proof.} \\
If $R_m := R(x_m , \mu)$ and
 $$ V^{(m, loc)}  (y) = R_m^2 V ( x_m + R_m y)\,, \hskip 2cm 
  U_{\ell} ^{(m, loc)}  (y)= R_m  U_{\ell}  ( x_m + R_m y)\,,  $$
 $$ A_k ^{(m, loc)}  (y) =   R_m  A_k  ( x_m + R_m y)  \,,
 \hskip 2cm  B_{jk}  ^{(m, loc)}  (y) =   R_m^2  B_{jk}   ( x_m + R_m y) \,.$$
If  $R_m \leq 1$, one verifies that, for any $(\Ab,W) \in \mathcal T (r,C_0)$, the corresponding pair $ ( \Ab ^{(m, loc)}, W^{(m,loc)})$ belongs to $\mathcal T (r,C_0)$.\\
  
 If  $R_m \leq 1/2$, we have  $\Phi ( x_m, R(x_m , \mu) ) = \mu$. 
 Applying Proposition \ref{VL-1} with $t= R_m = R (x_m , \mu)   \leq 1$, we have, if $|y|\leq 1$\,,
 $$ \Phi (x_m, R_m ) \leq C_2 \Phi ( x_m + R_m y , R_m )+ C_2 \leq C_2 \check m^{r , m, loc} (y)\,,
 $$
where   $\check m^{r , m, loc} (y)$ is the  function associated, as in (\ref{eq.mqx}), with the localized operator $ P_{{\Ab}^{m, loc},W^{m, loc}}  $ at the point $x_m$.  
Consequently we have $\mu \leq  C_2  \check m^{r , m, loc} (y) $ for all $y\in \mathbb R^d$ such that $|y|\leq 1$.\\
  
  By Theorem \ref{thcompresnew}, for all  $g\in C^{\infty} _0(\R^d)$ with support in 
$B(0, 1)$, we have :
$$  \mu^{\delta}  (1/C_2)^{\delta}
\Vert g \Vert \leq  C_1 ( \Vert P_{{\Ab}^{m, loc},W^{m, loc}}   g \Vert + \Vert g \Vert ) $$
where $C_1$ and $C_2$ depend only on  $C_0$. \\
Then one can find $\mu_0$ and $C_3$ with the same properties such that, for $\mu\geq \mu_0$:
 \begin{equation}\label{ELL}  \mu^{\delta}  \Vert g \Vert \leq  C_3 
  \Vert  P_{{\Ab}^{m, loc},W^{m, loc}} g \Vert\,. \end{equation}
 If  $f$ is supported in  $B( x_m , R(x_m , \mu))$, we apply (\ref{ELL})
to the function \break  $g(y) = f ( x_m + y R(x_m , \mu))$ and obtain for $\mu \geq \mu_0$
$$  
\frac {\mu ^{\delta } } { R(x_m , \mu)^2 } \Vert f \Vert
\leq C_3 \, \Vert P_{{\Ab},W} f \Vert  \,.$$
 Inequality  (\ref{interpol})  leads to  (\ref{EL}).

\subsection{End of the proof of Theorem \ref{t-princ}.} 

Let $u\in C_0^{\infty} (\R^d)$. For any  $\mu \geq 1$, we apply
 (\ref{phi-j-4}) and get,  distinguishing in the localization formula  the $x_m$ such that $R(x_m,\mu) >\frac 12$ from the terms such that $R(x_m,\mu) \leq \frac 12$, 
 $$  \int _{\R^d} \left [ \frac {|u(x)|^2 } {R(x , \mu)^4}   +
  \frac {|(D-\Ab) u(x)|^2 } {R(x , \mu)^2} \right ] dx
    \leq C  (\Vert u \Vert ^2 + \Vert (D- \Ab) u \Vert ^2 ) +R \,, $$
 $$ R =  C    \sum _{R(x_m , \mu) \leq 1/2}
  \frac { \Vert \varphi _m u  \Vert ^2 }  {R(x_m , \mu)^4}
  +  \frac { \Vert (D - \Ab)  (\varphi _m u)  \Vert ^2 }  {R(x_m , \mu)^2} \,.$$
Here we have also used  (\ref{phi-j-3})  for the control of  commutators. If  $\mu\geq \mu_0$ with $\mu_0$ large enough, for any  $m$ such that
 $R(x_m , \mu) \leq 1/2$, we apply Proposition  \ref{estim-loc}
 to the function $f = \varphi_m u$ and obtain:
 $$ R \leq C \mu^{- 2 \delta} \sum _{R(x_m , \mu) \leq \frac 12} \Vert P_{{\Ab},W} (\varphi_m u) \Vert ^2 $$
 $$ \leq  C \mu^{- 2\delta}  \Vert P_{{\Ab},W} u \Vert ^2 + C \mu^{- 2 \delta}
 \sum _m \Big [ \Vert \nabla  \varphi_m \cdot (\nabla -i\Ab) u \Vert ^2 +
  \Vert  u ( \Delta \varphi  _m) \Vert ^2 \Big ] \,.$$
  From (\ref{phi-j-3}), we deduce:
 $$R \leq  C \mu^{- 2 \delta}  \Vert P_{{\Ab},W} u \Vert ^2 + C \mu^{- 2 \delta}
 \int _{\R^d}\left [ \frac {|u(x)|^2 } {R(x , \mu)^4}   +
  \frac {| (D - \Ab)  u(x)|^2 } {R(x , \mu)^2} \right ] dx \,.$$
  There exists a possibly new $\mu_0$ such that, for  $\mu \geq \mu_0$, 
  $$ \int _{\R^d} \left [ \frac {|u(x)|^2 } {R(x , \mu)^4}   +
  \frac {|(D -\Ab)u(x)|^2 } {R(x , \mu)^2} \right ] dx
    \leq C  (\Vert u \Vert ^2 + \Vert (D -\Ab) u \Vert ^2 )
  + C \mu^{- 2 \delta}  \Vert P_{{\Ab},W} u \Vert ^2 \,.$$
  Using again  \eqref{interpol}, we get:
  $$ \int _{\R^d} \left [ \frac {|u(x)|^2 } {R(x , \mu)^4}   +
  \frac {|(D - \Ab ) u(x)|^2 } {R(x , \mu)^2} \right ] dx
    \leq C  \Vert u \Vert ^2
  + C (1 +  \mu^{- 2 \delta})  \Vert P_{{\Ab},W} u \Vert ^2 \,.$$
  Theorem  \ref{t-princ} follows since 
  $ \Phi (x , R(x , \mu) ) \leq \mu$ and consequently 
  $$ R(x , \mu) \sum _{\ell} |U_{\ell} (x)| + R(x , \mu)^2 
  \sum _{j <k} |B_{jk}  (x)| +  R(x , \mu)^2 |V(x)|  \leq \mu \,.$$
  \begin{remark}
  The proof of Theorem \ref{t-princ} does not give directly the maximal estimates as described in the introduction. But the control of the theorem reduces this question to the analysis 
   for a selfadjoint operator which could be either the magnetic Laplacian ($W=0$), or the magnetic Schr\"odinger operator $P_{\Ab,U}$ or the operator $P_{\Ab,U+\sqrt{1+V^2}}$. For these operators one can for example use Shen's  Theorem \ref{thSh}  (at least when $d\geq 3$, but see Remark \ref{shend2} for $d=2$). Note that to be in $\mathcal T (r,C_0)$ is usually not enough except if $\Ab$ is a polynomial. In this case, one can use the nilpotent approach for getting the maximal regularity of the magnetic Laplacian. Otherwise, one can for example prove the complete maximal estimates
    under the condition that $(\Ab,W=0)$ belongs to $\mathcal T (r',C_0')$. 
    \end{remark}
    
   {\bf Acknowledgements}\\
    The authors would like to thank J. Camus for the transmission of \cite{Gu1} and Z. Shen for an enlightning clarification \cite{Sh3}. These results were announced at the centennial Kato conference in Tokyo (September 2017). The first author would like the RIMS Kyoto, Y. Nakamura and K. Yajima for their support.\\

\end{document}